\documentclass[11pt,tightenlines,eqsecnum,floats,aps,nofootinbib,prd,showpacs]{revtex4}
\usepackage{hyperref}
\usepackage{amsmath,amssymb,amsfonts,amsthm,amscd}
\usepackage{graphicx}
\usepackage{enumerate}
\usepackage{colordvi}
\usepackage{units}
\usepackage{epsfig}
\usepackage{natbib}
\usepackage{enumerate}
\usepackage{colordvi}
\usepackage{multirow}
\usepackage{afterpage}
\usepackage{pdflscape}
\usepackage{color}
\def\be{\begin{equation}}
\def\ee{\end{equation}}
\def\ba{\begin{eqnarray}}
\def\ea{\end{eqnarray}}

\begin{document}

\title{Solving the cosmological entropy issue with a Higgs dilaton}

\author{David Sloan}
\email{david.sloan@physics.ox.ac.uk}
\affiliation{Beecroft Institute of Particle Astrophysics and Cosmology, Department of Physics,
University of Oxford, Denys Wilkinson Building, 1 Keble Road, Oxford, OX1 3RH, UK}
\author{George Ellis}
\email{george.ellis@maths.uct.ac.za}
\affiliation{Mathematics Department, University of Cape Town, Rondebosch, Cape Town 7701, South Africa}

\vspace{0.1in}

\begin{abstract}

\textbf{Abstract:} \textit{Current cosmological models require the universe to be in a very smooth initial state before the onset of inflation, a situation to which Penrose ascribes a vanishingly small probability, leading to his proposal of a Conformal Cyclic Cosmology. We present an alternative paradigm, in which the Higgs plays the role of dilaton and resolves this problem by weakening gravity at very early times, thus providing a form of inflation that is compatible with observations and in which the inflaton is solidly related to tested particle physics.} 
 
\end{abstract}

\maketitle

\section{Introduction}
This paper proposes a unified solution to two issues of major import in current cosmology. First, what is the nature of the inflaton? Second, why is the early universe in a very special state that allows inflation to start, when that is a highly improbable situation? We propose solving both  simultaneously by using a Higgs dilaton as the inflaton, but coupled in such a way that gravity is essentially turned off in the very early universe. This unifies proposals made by others into a coherent whole that is well worth exploring further.\\

\textbf{The nature of the inflaton} The theory of inflation is a huge success, being able to explain cosmological observations to high precision \cite{Planck1,Planck}. However it has a major defect: we have no solid basis for stating what the inflaton is. A great many models have been put forward \cite{encyc}, with some fitting the observations better than others; but most have no solid basis in established physics. There is however one exception: Higgs inflation is possible \cite{Bezrukhov}\citealp{Garcia} and fits the observations very well \cite{encyc}. This is the one and only case in which the inflaton is related to tried and tested physics \cite{EllUzan}, because the Higgs has indeed been observed at the LHC. Because its properties underlie key features of the standard model of particle physics, if the Higgs were to be the inflaton, one would have one of the most awesome unifications imaginable in physics: the same particle is responsible both for mass at the microscale,  and for the dynamics of the very early universe, and hence controls the seeds of structure formation  at macro scales. This is therefore a proposal  that should be seriously pursued to see if it might work \cite{EllUzan}. It is the one possibility for an inflaton solidly tied in to the standard model of particle physics.\\  

\textbf{The fine tuned initial state} The second point is that while it is often stated that inflation solves the flatness and horizon problems in the early universe, that is not in fact the case, as Penrose has pointed out in various writings that are summarised in Chapter 3 of \textit{Fashion, Faith, and Fantasy} \cite{Penrose}. The essential point is that the maximum entropy of a given amount of matter is attained by collapsing it into a black hole. By contrast, ``\textit{the Big Bang was an event of extraordinarily low entropy ... the gravitational degrees of freedom were completely suppressed}''  (\cite{Penrose}:258). Penrose estimates the extraordinary precision that was involved in setting the initial state of the universe as $10^{-10^{123}}$ (\cite{Penrose}:275). This situation is not made clear by standard inflationary studies because they consider only  perturbed Robertson-Walker geometries.\\

Penrose proposes to solve this problem by invoking a \textit{Conformal Cyclic Cosmology} (\cite{Penrose}: \S4.3). This is a creative idea but the mechanisms involved in realizing such a proposal are not all clear. \\

\textbf{Turning off gravity at very high densities} By contrast, we propose to invoke a mechanism proposed by Greene \textit{et al} \cite{Greene}: namely that the strength of the gravitational force dies away at very high energies. In that case the collapse to black holes that causes the problems identified by Penrose will not occur: thermal forces will be able to overcome gravitational forces in the very early universe and prevent collapse to blackholes.   The gravitational degrees of freedom will not be suppressed, they will simply not dominate the degrees of freedom of ordinary matter as in the standard case considered by Penrose. Given enough time, thermal processes will result in a uniform state of matter, which will allow the geometry to also be smooth. There is an issue here as to whether there will be a sufficient time available for such equilibriation to take place; that will not always be the case. However if we assume the universe starts off in a most probable state, that will be a state of maximal entropy, which will indeed be smooth as in the case where gravity is completely switched off. This may be taken as a reasonable initial assumption, expressing the idea that the starting configuration is the most probable one.\\

\textbf{Putting them together} Thus we propose to combine these ideas: we will use a Higgs inflaton in a context where gravity is turned off at high densities by the kinds of couplings considered by Greene et al \cite{Greene}.\footnote{A somewhat similar proposal was made by Alexander \textit{et al} \cite{Barrowetal} but they did not relate it to the gravitational entropy problem identified by Penrose. In any case the inflaton dynamics involved in that case does not work (Section \ref{sec:frame_trans}). } 
In physical terms, we make Newton's constant dynamical \textit{a la} Jordan-Brans-Dicke but with the Higgs playing the role of the coupling. We aim to show this is then a viable cosmological theory that embodies the unifications identified above.\\

\textbf{Black hole entropy} The Bekenstein-Hawking entropy is determined by the surface area of a black hole as would be measured in flat space, and fundamental constants, and is given:
\be S_{BH} = \frac{c^3A}{4G\hbar} \label{BH} \ee
In the case of a Schwarzschild black hole, $A=16 \pi \left(\frac{GM}{c^2} \right)^2$ and hence we find that we should expect that the gravitational entropy of black holes is linearly dependent on the strength of the effective Newton constant. Note that we are thus broadly in keeping with the second law of thermodynamics only if $G$ is increasing.\\

\textbf{Naturalness} Proposals for Higgs inflation have sometimes been criticized as beung ``unnatural'', as they require very large values of the coupling parameter between  the Higgs field and gravity. However Hossenfelder \cite{Hossen} has trenchantly pointed out that such `naturalness' criteria may be very misleading, and should not in general be taken as hard guidelines for physical theories. In our case we adopt the same position: given the other advantages flowing from the proposed unification, we do not see that that any naturalness objection destroys our proposal. \\

\textbf{A basis in quantum gravity theory?} We do not at this time have any proposal as to how the effective theory we put forward might have a deeper foundation in an underlying more fundamental theory of gravity. This would obviously be desirable. We take our choice to be justified by the unification of different aspects of cosmology achieved, as set out above.\\

\textbf{This paper} Section \ref{sec:Foundations} deals with foundations. 
Section \ref{sec:small} considers the case where the field starts off at small values, and Section \ref{sec:large} the case where the field starts off at large values.  Section \ref{sec:frame transformation} deals with the effect of frame transformations, and Section \ref{sec:further} considers issues that arise.

\section{Foundations}\label{sec:Foundations}

We begin considering 
a scalar field $\phi$ non-minimally coupled to gravity through some function $F(\phi)$ which multiplies the scalar curvature $R$. Then the action is 
\be S = \int \sqrt{g} \left( \frac{F(\phi)}{6} R + \frac{1}{2} \partial_a \phi \partial^a \phi - V(\phi) \right) \label{eq:action}\ee
where $V(\phi)$ is the potential. We have chosen the factor of 6 in the normalisation of the function $F$ to make algebra more convenient when specializing this to the case of a Robertson-Walker geometry. As an aside, we can here make a direct comparison with dilatonic gravitational actions inspired by Brans-Dicke with dilaton $\Xi$:
\be S_D = \int \sqrt{-g} \left( [ \Xi R - \frac{\omega(\Xi)}{\Xi} \frac{1}{2} \partial_\mu \Xi \partial^\mu \Xi] - V_D [\Xi] \right)\label{eq:B_D} \ee
Here we see that the two are indeed equivalent under the identification $6 \Xi=F$, $\omega = 6 \frac{F}{ F'^2}$ and $V = V_D \circ 6F^{-1}$. Working directly with the dilatonic action is unwieldy given our cosmological aspirations, hence we will continue from the original action. \\

\textbf{Cosmology} Let us now restrict ourselves to a homogeneous, isotropic spacetime with curvature $k = \pm 1, 0$. Thus the partial derivatives are simply time derivatives (up to a choice of lapse).
 The Ricci scalar in the case of a Robertson-Walker geometry is:
\be R = 6 \left( \frac {\ddot{a}}{a} + \frac{\dot{a}^2}{a^2} + \frac{k}{a^2} \right) \ee
where $a(t)$ is the scale factor and $k = 
\{\pm 1, 0\}$ the normalised spatial curvature. 
 Since the action contains second derivatives of the fields, we integrate by parts and recover (up to boundary terms) a Lagrangian which closely resembles the familiar case of minimally coupled matter, with an extra term arising from the variation of $F$,  with $F$ implicitly dependent upon $\phi$:
\be \mathcal{L} = a^3 \left(-F \left(\frac{\dot{a}^2}{a^2} + \frac{\dot{a}F'}{aF} \dot{\phi} - \frac{k}{a^2}\right) +\frac{\dot{\phi}}{2} - V \right) \ee
wherein primes denote derivates with respect to the field $\phi$. 

The dependence of $F$ on the field $\phi$ leads to modifications of the momenta conjugate to $\phi$ and $a$ from their usual forms. Varying the Lagrangian with respect to the velocities of each of these in turn yields:
\be P_\phi = a^3 \dot{\phi} - F' a^2 \dot{a},\,\, \quad P_a = -F' \dot{\phi} a^2 +2F\dot{a} a = -\frac{d}{dt} (Fa^2). \ee
Thus we can find the Hamiltonian (the Noether charge associated with time translation), which determines the Friedmann equation from the Einstein-Hilbert action:\footnote{The cosmological equations are given in the Appendix.}
\be \mathcal{H} = -a^3 F \left(H^2 + \frac{F'}{F} \dot{\phi} H + \frac{k}{a^2}\right) + a^3 \left( \frac{\dot{\phi}^2}{2} + V \right) :=0 \ee
wherein we have introduced the standard terminology $H=\dot{a}/a$ for the Hubble parameter. We note that in the case of the reduction  $F' \rightarrow 0$ this reproduces the usual Friedmann equation, however otherwise we have a new term introduced by the variation of $F$.  The extra term breaks the usual relationship between expansion and energy density, thus we should expect to see more interesting dynamics away from stationary points of $F$, as we will have to solve a quadratic equation for $H$. Introducing \be \rho = \frac{\dot{\phi}^2}{2} + V\ee in the case of the scalar field (in general $\rho$ represents the energy density of any matter present and minimally coupled to gravity) we see:
\be H = -\frac{F' \dot{\phi}}{2F} \pm \sqrt {\frac{F'^2 \dot{\phi}^2}{4F^2} + \frac{\rho}{F} - \frac{k}{a^2}} \label{Friedmann} \ee
Since the additional term under the square root is positive for $F>0$ such functions would not introduce qualitatively new features (such as bounces in the $k=0,-1$ cases) through the role of $F$. We also retain two branches to cosmological solutions, one expanding $(H>0)$, the other contracting $(H<0)$. 

The Euler-Lagrange equations for our system give the dynamics of the fields. The motion of the scalar field is the usual Klein-Gordon equation modified by a gravitational source term proportional to $F'$:
\be \ddot{\phi} + 3H \dot{\phi} +V' = F'(\dot{H}+2H^2 + \frac{k}{a^2}) \label{KG} \ee
and the usual Raychaudhuri equation is replaced by
\be 2F\dot{H} + 3FH^2 +F\frac{k}{a^2} +2F'\dot{\phi}H+F'\ddot{\phi} +F''\dot{\phi}^2 =3(V-\frac{\dot{\phi}^2}{2}) \label{Raychaudhuri}. \ee
These equations each reduce to their regular forms in the case $F', F'' \rightarrow 0$, but have corrections away from such points. We can combine them to find a more informative version of the Klein-Gordon equation by eliminating the gravitational source term. Using \ref{Raychaudhuri} to eliminate terms in $\dot{H}$ and \ref{Friedmann} to deal with the remainder we arrive at
\be \ddot{\phi} + 3H\dot{\phi} + \frac{V' +\frac{F'}{F} \left((1-F'')\frac{\dot{\phi}^2}{2}-2V\right)}{1+\frac{F'^2}{2F}} \,=\, 0\label{KG2}. \ee
An important feature  is that it reveals that the stationary point $(\ddot{\phi}=\dot{\phi}=0)$ of the scalar field may have shifted - it is no longer necessarily at the minimum of the potential but rather is determined by:
\be \frac{V'}{V} = 2\frac{F'}{F} \label{newmin} \ee
Here we note that depending of the functional forms of $V$ and $F$ there could be many realizations of the above condition, or none. If we find there to be such a condition away from the minimum of the potential, this appears as a new term in the effective matter energy density of the system; if viewed through the lens of minimally coupled GR, it is an effective cosmological constant, and potentially has interesting ramifications for reheating. 
 If this does not occur, the minimum will be at the Higgs VEV. In either case, at the stationary point $F$ becomes fixed. Thereafter the system will be described by GR with a fixed Newton constant, whose value is set by the inverse of F at this stationary point.

\section{Small Initial Field}\label{sec:small}

We first consider the case where our scalar field begins at small field values, close to the origin. This of course raises issues of naturalness - it would appear that the field would need fine-tuning for this to occur in nature. However, we shall see that there are certain situations in which this can be alleviated. 
Specifically we take:
\be F= A^2 \exp \left[-\left(\frac{\phi}{S}\right)^n\right], \quad \quad \quad  V = (\Phi^2- \phi^2)^2 \label{eq:Higgs_11}\ee
with $A$,  $S$, and $\Phi$ constants and $n$ a postive integer and will include other minimally coupled matter, labelled by its energy density  $\rho$. We will choose $S^2<<\Phi^2$ so that the dynamics of the Higgs potential is close to that of the standard model when we approach the minimum of the potential. \\

\textbf{Gaussian $F$} The first thing to notice is that in the case that $F$ is Gaussian (n=2) we find that $\phi=0$ becomes a stable minimum of the system. To see this consider the Klein-Gordon equation in this setup. Linearly perturbing $\phi$, to order $\epsilon$ in both $\phi$ and $\dot{\phi}$ we see that:
\be \ddot{\epsilon} + 3\sqrt{\rho + \Phi^4} \dot{\epsilon} + \left(\frac{\rho}{S^2} + \frac{4\Phi^4}{S^2} - 4\Phi^2 \right) \epsilon \,\,=0  \ee
and our condition on $S^2$ means that the coefficient of $\epsilon$ is positive. In such a case we could relax our tuning of the system somewhat, as small perturbations about the hilltop in $F$ and $V$ would not lead to run-away dynamics. Any initial condition with $\phi_0^2 < \Phi^2 - S^2$ will tend toward $\phi=0$, and settle there classically. However, this condition holds indefinitely, and as such it would not be possible for the particle to escape from this setup, unless it were to tunnel. That would be the only way to end inflation.

At this point it is useful to note that this system will have five stationary points: $\phi=0, \phi=\pm \Phi$ are stable, $\phi=\pm \sqrt{\Phi^2-S^2}$ are unstable. To see this consider whether the Klein Gordon equation with $\dot{\phi}=0=\ddot{\phi}$ can be satisfied by solutions to equation \ref{newmin}, which in this case implies the conditions for unstable equilibria. Likewise $\phi=\Phi$ is stationary. To see stability, we can expand our solutions to first order in $\phi,\dot{\phi}$ about these points. Setting $\phi=\sqrt{\Phi^2-S^2}+\epsilon$ we see
\be \ddot{\epsilon}+3H\epsilon - \frac{8H(\Phi^2-S^2)}{H+2(\Phi^2-S^2)} \,\, =0 \ee
hence these are unstable equilibria, as the coefficient of $\epsilon$ is negative. Now let us consider the minima of the Higgs potential: setting $\phi=\Phi+\epsilon$ we find
\be \ddot{\epsilon} + \frac{ 8\ S^4 \Phi^2}{S^4+2 \Phi^2 A^2 \exp[-\Phi^2/S^2]} \epsilon \,\,=0 \ee
where the coefficient of $\epsilon$ is positive and hence we have a stable equilibrium. At this point the effective Planck mass is $M_p^2 = A^2 \exp[-\Phi^2/S^2]$. Fixing this at this point to match with observations made in the late universe, we see that if the field can move away from the equilibrium point at the origin out to the Higgs minimum by a tunnelling process, the effective Newton constant will have fallen by a factor of $\exp[\Phi^2/S^2]$. 

In order for the classical motion to no longer move the field towards the hilltop the tunnelling would have to take the particle beyond $\phi=\sqrt{\Phi^2-S^2}$ - the next (unstable) equilibrium point of the system - and close to the minimum of the potential. From this point onwards the dynamics could reproduce that of slow-roll inflation: The slow roll parameters\footnote{We here use the geometric version of the slow-roll condition rather than the potential form, as it is $H$ not $V$ which determines the behaviour of perturbations, and our dynamics is not simply dependent on $V$.} are $\epsilon_H = \frac{-H'}{H^2}$ and $\eta_H = \frac{H''}{H'H}$, which to first order about this point can be expressed in terms of the field difference from the point, $\delta=\phi-\sqrt{\Phi^2 - S^2}$: 
\be \epsilon_H \approx \frac{8A \Phi^3 \delta}{S^4 \exp[\frac{\Phi^2}{2S^2}]} = \frac{8M_p \Phi^3}{S^4} \delta ,\quad\quad\quad \eta_H \approx \frac{24A \Phi^3 \delta}{S^4 \exp[\frac{\Phi^2}{2S^2}]} = \frac{24M_p \Phi^3}{S^4} \delta  \ee
at which points the Hubble parameter takes the value $H=\frac{S^2}{A} \exp[\frac{\Phi^2}{2S^2}] = \frac{S^2}{M_p}$. Thus we see that we have simply replaced one issue of fine-tuning with another; the system would have to tunnel incredibly close to the unstable equilibrium point to begin a slow-roll inflation which was compatible with observations. We therefore turn our attention to other possible choices of $F$. \\

\textbf{Even powers of $\phi$} The next logical one is given by setting $n=4$. Unfortunately this, and all higher (even) powers also suffer from having multiple stable minima that are away from the minimum of the Higgs potential. It is, of course, possible to tune such minima so that one of them is compatible with the cosmological constant, however doing so would be essentially no less arbitrary than adding a constant to the potential to recover the same effect. An interesting case in which this occurs due to a more natural mechanism created by the Higgs vacuum being dynamic is presented in \cite{Davies}.\\

\textbf{Free choice of $F$} Let us now consider whether a small initial field choice can ever resolve the problem of gravitational entropy, with the function $F$ kept largely free. In order to have the field begin at small values we require that at least the point $\phi=0$ be an equilibrium point of the system. For our argument we do not here need to require stability at this point. From the Klein-Gordon equation we see that $\phi=0$ being an equilibrium requires that $F'=0$ at this point. Our second requirement is that we want to avoid there being a stationary point of the system for $\phi \in [0, \alpha \Phi]$ for some positive number $\alpha$ so that our cosmology does not enter a deSitter phase that is incompatible with observations. From the current value of the cosmological constant and the potential $V$ we see that this requirement is that $\alpha \approx (1-10^{-37})$. Our requirement that there be no stationary points means that equation \ref{newmin} cannot be satisfied on this interval. Hence:
\be 2(\log F)' < (\log V') \rightarrow \frac{F(0)}{F(\alpha \Phi)} < \sqrt{\frac{V(0)}{V(\alpha \Phi)}} = \frac{\Phi^2}{\sqrt{\Lambda}} \approx 10^{37} \ee

Let us summarise this result: If we want to have the effective Newton constant $G$ be a function of the Higgs field such that it does not generate a cosmological constant which is significantly larger than the one currently observed 
  and the field begins its dynamics at the origin, then the maximum ratio of the value of $G$ today to its early value is approximately $10^{37}$. Whilst this is a very large number, it is not enough to overcome the criticisms that Penrose has levelled about the fine-tuning of the initial state: the initial entropy per baryon would be dropped by $10^{37}$, which is not enough to overcome the factor of $10^{123}$ that Penrose calculates (he calculates the \textit{probability} of being in 
this state to be $10^{{10}^{123}}$ 
  and the  \textit{entropy} is the log of this, hence $10^{123}$). Starting near the origin appears to fail entirely 
  unless one is willing to have \textit{very} fine-tuned quantum tunnelling out of such a phase and have $F$ extremely high at the origin. %

Note further that this is a very liberal estimate on the upper bound for this ratio; if we were to make the more conservative assumption that the dynamics of the Higgs field was unaffected in a larger region about its minimum we would obtain a much lower bound. For example, insisting that the effective Newton constant is within a few percent of its current value for $\phi>0.99 \Phi$ would restrict this ratio to being in the region of 50.\\

\textbf{In conclusion:} if we don't want to require tunnelling nor introduce a new cosmological constant this really rules out the small field case as a solution to the entropy problem pointed out by Penrose. A (stable) stationary point of the scalar field away from the minimum of the potential introduces a cosmological constant, and the classical dynamics comes to a rest at this point. We want to rule that out if it would give an effective cosmological constant orders of magnitude bigger than our own today. This means that there's a condition relating the ratio of the potentials today and at the origin,  and the ratio of the function $F$ today and at the origin. This in turn puts an upper bound on the ratio of G today to it's minimum value, which, even in the most liberal case, is about $10^{37}$, and in much more realistic cases is more like 100 - vastly less than needed to solve the Penrose problem. The only way in which this scenario can be avoided is to very precisely define $F$ such that $G$ becomes equal to its current value very suddenly, requiring a degree of fine-tuning on at least the same order as the problem that Penrose identifies.     We are therefore motivated to move away from looking for situations in which the field begins at small values and investigate the case of a large initial field. 

 \section{Large Initial Field}\label{sec:large}
 
An alternative way in which we can realize the dynamics we seek is to consider beginning the scalar field at high (positive) field values. We note that in such cases $F$ should be monotonically increasing towards high field values so that equation (\ref{BH}) obeys the second law of thermodynamics. To examine this possibility, consider the dynamics of a field that begins at rest at a high field value. From equation \ref{KG2} we see that the acceleration of the field is proportional to $2V\frac{F'}{F} -V'$. Thus we see that for this acceleration to be towards the minimum of the potential we need that asymptotically $F^2 \leq V$, which in the case of our Higgs potential means that $F$ cannot asymptote to a function which grows faster than $\phi^2$. \\

\textbf{Quadratic $F$} The case in which $F \propto \phi^2$ is that examined in many cases of Higgs inflation (see e.g. Shaposhnikov \cite{Bezrukhov}) and this term means the acceleration of the field tends to zero at large field values, giving rise to the slow-roll effect, as needed for inflation. This is usually established in the Einstein frame for convenience of calculation, but here we retain the Jordan frame as we will examine more general scenarios. 

In recent work it was suggested that this alteration of the effective Newton constant could be responsible for `turning off' gravity at high energy densities \cite{Barrowetal}. Here we demonstrate why this limit is still singular in our case, and in section \ref{sec:frame transformation} we give a general argument as to why this cannot be achieved. The Hubble parameter is determined by equation \ref{Friedmann}, and thus grows at high field values. Since both kinetic and potential energy contribute to $H$, and both are positive, the Hubble parameter at a given value of $\phi$ is greater than the value it would take if the kinetic energy were zero:
\be H > \sqrt{\frac{V}{F}} > V^{1/4} \propto \phi \label{HFV} \ee
by the same consideration in the case of a Higgs potential. Thus we see from running our dynamics backwards that in such cases the cosmological model begins with an initial singularity ($H \rightarrow \infty$ at $\phi \rightarrow \infty$). We shall show that this behaviour is universal in such models. The quadratic case is the fastest rate of growth of $F$ in which the field is driven to the minimum of its potential (see section \ref{sec:frame_trans}) but for our purposes the growth need not be quadratic. We thus further split the discussion into two cases; the first case to consider is where $F$ is unbounded from above (but grows slower than $\phi^2$), the second where $F$ asymptotes to a finite constant. \\

\begin{figure}
\begin{center}
\includegraphics[width=0.7\textwidth]{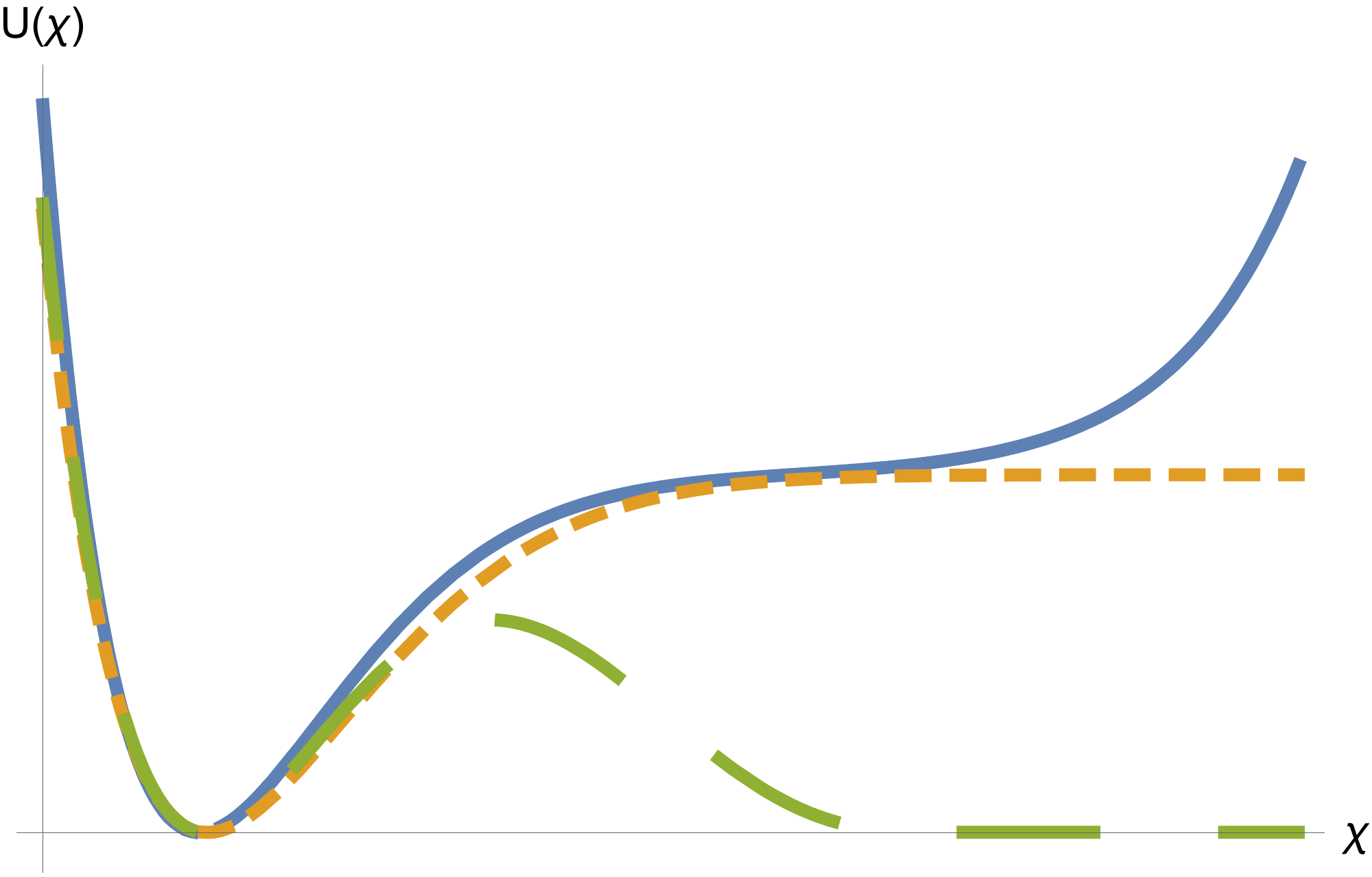}
\caption{\textit{Potential in $\chi$ for three different choices for $F$ corresponding to the usual quadratic (orange, short dashed), a quartic (green, long dashed) and $F= \frac{\phi^2}{A^2 + \phi}$ (blue, solid). We see that in the latter two cases we recover the plateau for Higgs inflation near the minimum of the potential, and in the first case the field would roll off to infinity as described.}}\label{LargeFields}
\end{center}
\end{figure}

\textbf{If $F$ is unbounded above} let us consider the following as an exemplar function:
\be F= \frac{A^2 \phi^2}{(\phi+\xi^2)}\label{eq:phi_square} \ee
with $A$, $\xi$ constants, and $V$ is given by (\ref{eq:Higgs_11}). 
We  choose $\xi^2 >> \Phi^2$, and could use the modulus of the numerator to make an even function if we wish to consider negative field values. In such a model we see the asymptotic behaviour required - $F$ becomes linear in $\phi$ at high field values, and is quadratic for $\phi < \xi$. We note that the condition for finding a non-trivial minimum was given in equation (\ref{newmin}). Here we show that this cannot be realized: 
\be \frac{V'}{V}-2\frac{F'}{F} = \frac{2 \xi^2 \Phi^2 +\Phi^2 \phi +6 \xi^2 \phi^2+7 \phi^3}{2 \phi  \left(\xi^2+\phi \right) \left(\phi^2-\Phi^2\right)} \label{eq:function}\ee
which is never zero (or singular) away from $\phi=\Phi$. Therefore the complete dynamics of such a system closely matches those we want: the field begins at large  field values where the effective Newton constant is almost zero, evolves through a phase in which it undergoes Higgs inflation, and eventually comes to rest at the minimum of the Higgs potential. The behaviour of this system is analysed in the Einstein frame in section \ref{sec:frame transformation}, with the effective potential shown in figure \ref{LargeFields}. Thus this model resembles the standard Higgs inflation case \cite{Bezrukhov} and so can fit the Planck data \cite{encyc}. The current value of the Planck mass is (for $\xi >> \Phi$) given by $M_p = A \Phi/\xi$, whereas at high values of the scalar field (i.e. in the early universe) it tends towards $A \phi$, and hence tends to infinity ($G \rightarrow 0$) at the initial singularity, thus removing the initial entropy problem. \\

Overall: this is a case that works.\\

\textbf{$F$ asymptotes to a finite constant} In the second case we mirror the ideas of, for example T-models or $\alpha$-attractors in which inflation is reproduced by a scalar field with a potential that tends towards a plateau at high field values. In these situations slow roll inflation is achieved since the potential is sufficiently flat far from the origin, and finding the inflaton here should not be considered unnatural as there is always an infinity of parameter space with higher field values. As an example of this, we will choose the function 
\be \label{eq:tahnh_square} F=M_o^2 + A^2 \tanh^2(\phi/S),\ee 
with $M_0$, $A$, and $S$ constants, where we shall choose $S$ to be considerably larger than the Higgs VEV. Note that there are a plethora of functions we could have used in this instance each of which although quantitatively distinct, we should expect to have qualitatively similar features. Our choice is motivated simply by the quadratic nature of the function about zero and the good approximation as such around the Higgs VEV, matching the usual Higgs inflation coupling, and its asymptotic plateau. In such a case, since the terms $F'^2/F$ and $F' V/F$ asymptote to zero, we see that in the large field limit we should expect to see the dynamics of the scalar field match those of general relativity, albeit with a significantly reduced Newton's constant. However, once the field exits the plateau phase its dynamics tend towards those of Higgs inflation (see e.g. \cite{Bezrukhov}) which agree with the Planck observations \cite{encyc}.
 
Let us note a few features of such choices of $F$. First we can show that the asymptotic behaviour of the system is indeed that the field comes to rest at the minimum of the Higgs potential. Again, the condition for a stationary point away from the Higgs minimum cannot be realized:
 \be \frac{V'}{V}-2\frac{F'}{F} = \frac{\phi^2 - \Phi^2}{4 \phi} - S \cosh^2(\frac{\phi}{S}) \coth(\frac{\phi}{S})(M_o(1+A)\tanh(\frac{\phi}{S})) \ee
First we consider the case $\phi>0$. Finding a Laurent series for $V'/V-2F'/F$ reveals that this can be expressed as a polynomial in $\phi$, the coefficients of each term being negative. Hence the sign of the function is fixed. Now since the total expression is odd, a parallel argument runs for $\phi<0$. Therefore the only possible points at which the field can be at rest are at the origin (which is an unstable equilibrium point, and thus would only hold if the field were always there) or at the minima of the Higgs potential (stable equilibria). Hence the field will eventually come to rest at the Higgs VEV. For $S >> \Phi$ we find that this gives an effective Planck mass of 
\be M_p^2 = M_o^2(1 + A^2 \tanh^2(\frac{\Phi}{S})) \approx M_o^2  \ee
whereas at high field values, the effective Planck mass is determined on the plateau of $F$:
\be M_p^2 \approx M_o^2(1 + A^2) \approx A^2 M_o^2 \ee
Thus we see that if we want to alleviate the initial fine-tuning problem we 
require that $A$ is of the order of $10^{123}$, thus reducing $G$, and hence the entropy count of primordial black holes, by this same factor.\\

In this bounded-above case, you do need a large number somewhere to make the potential high enough to overcome the $10^{123}$ factor in $G$
. It does look like an unnaturalness in terms of having such a large number instead of one around unity; but a number even larger than this would still work, so it's not fine tuning in the sense of needing a rather specific value. Also  on using a log prior \cite{Jeffries}
  it's only tuned to around 1 part in 100. If we in any case adopt the view that naturalness of such parameters is not a major issue \cite{Hossen}, this case works too. 
However perhaps the unbounded-above case mentioned above is more appealing 
 as you only need to introduce one new term ($\xi$) which is bounded to be an order of magnitude above the Higgs minimum. 
  
 \section{Frame Transformation and Potential}\label{sec:frame transformation}\label{sec:frame_trans}
 
 Our analysis thus far has been carried out in the Jordan frame. However, if we want to make comparisons with inflation, it is often useful to transform to the Einstein frame. Here we will follow Kaiser (94), and note that by making a conformal transformation of the metric ($\hat{g}_{\mu \nu} = \Omega^2 g_{\mu \nu}$) we can rewrite the non-minimally coupled field as a scalar field with a transformed potential. The choice $\Omega^2 = F/6$ renders the action
  \be S=\int \sqrt{\hat{g}} \hat{R} + \frac{ \partial_\mu \chi \partial^\mu \chi }{2} - U(\chi)  \ee
 wherein $\hat{R}$ is the Ricci scalar of the metric $\hat{g}$, and $\chi$ is obtained from solving
 \be \frac{d \chi}{d\phi} = \sqrt{\frac{F+3(F')^2}{4F^2}} \ee
 with the transformed potential 
 \be U(\chi) = \frac{V(\phi)}{4F^2} \ee
\textbf{Unbounded-above case} Let us return our attention to the unbounded large field case  where the function $F$ is that of (\ref{eq:function}):  
$F = \frac{A^2 \phi^2}{\xi^2 + \phi}
$. To analyse the choices in the Einstein frame, we first note that the potential $U(\chi)$ can be expressed as
\be U = \frac{V}{4F^2} = \frac{(\Phi^2 - \phi^2)^2}{4A^2 \phi^4} (\xi^2 + \phi)^2 \ee
and hence it becomes apparent that we should split our analysis into three regimes 
 corresponding to the initial large field, the region in which $A$ is most relevant, and the behaviour around the minimum of the Higgs potential: $\phi >> \xi$, $\Phi < \phi  <<\xi$ and $\phi \approx \Phi$ respectively. In the first of these regimes 
  we see that $F$ is essentially linear, and hence $\chi \propto \sqrt{x}$, and thus $U(\chi) \propto \chi^4$, and our field will begin its descent down the potential. In the latter two regimes 
   $\chi \propto \sqrt{\phi}$ and hence at first $U(\chi) \approx \xi^4 + \frac{2\xi^2}{\chi^2}$, during which phase the potential is approximately a plateau, and the dynamics will match that of Higgs inflation, since $F \approx \frac{A^2\phi^2}{\xi^2}$ and thus our action would match that of Higgs inflation. In the final phase we can expand $U$ about $\Phi^2$, and we see that we once again recover the Higgs potential; 
\be U = \frac{(\Phi+\xi^2)^2}{4A^2 \Phi^4} \left(\chi-\Phi^2\right)^2 + O(\chi-\Phi^2)^3 \label{eq:Higgs_pot}\ee

Equation (\ref{eq:Higgs_pot}) refers to 
 the unbounded-from-above potential, but this time in the Einstein frame rather than the Jordan frame. This is an example that works. The point here is that you can see why, for example, $F>\phi^2$ doesn't work, and the nice plateau form of the potentials in this setup.\\

\textbf{What does not work} Consider the case in which $F$ grows faster than $\phi^2$ for large field values. To make this explicit, consider $F \propto \phi^n$, where we shall later set $n>2$. Then when we transform to the Einstein frame, we find 
\be \Omega^2 \propto \phi^n , \quad \chi \propto \log \phi + O(\phi^{-2}) \ee
which together tells us that at large field values $\Omega^2 \propto e^{\frac{2\chi}{\sqrt{3}}}$ and thus we see that the transformed potential in the Einstein frame is given by 
\be  \label{Ularge} U(\chi) \propto \frac{(\Phi^2-e^{\frac{4 \chi}{\sqrt{3}n}})^2}{e^{\frac{4\chi}{\sqrt{3}}}} \rightarrow e^{\frac{4\chi}{\sqrt{3}n}(2-n) }. \ee
Hence if $n>2$ the potential is decreasing away from the origin, and the field will roll away to infinity in such cases. Thus our analysis in the Einstein frame matches that in the Jordan frame for these choices of high initial field values. In other words, if $F$ asymptotically grows faster than $\phi^2$ our cosmological solutions will push $\phi \rightarrow \infty$ forever, and we will recover neither inflation nor standard model physics. In fact this result only relies on the fact that $V$ must grow faster than $F^2$ to have a monotonically non-decreasing potential in $\chi$, since $\chi$ is by design a monotonic function of $\phi$. Together with the result of equation \ref{HFV} which tells us that for generic potentials $V$ and functions $F$, $H \propto \sqrt{FU}$, we see that if both $F$ and $U$ are non-decreasing functions the only way to avoid a $H \rightarrow \infty$ singularity at large $\chi$ would be if both were bounded from above.

 Let us consider that scenario. If we were to modify the Higgs potential such that it asymptotes to a constant we would still find the singularity unavoidable. Suppose $V=V_\infty$ and $F=F_\infty$ for $\phi > \phi_m$ say for some finite values of $V_\infty$ and $F_\infty$. Then the Klein-Gordon equation \ref{KG2} reduces to a (somewhat rescaled) version of the usual inflaton dynamics on a flat potential:
\be \ddot{\phi} + 3\dot{\phi} \sqrt{\frac{\dot{\phi}^2}{2F_\infty} + \frac{V_\infty}{F_\infty}} = 0 \ee
and thus the kinetic energy of the inflaton tends to infinity, taking the Hubble parameter with it. Hence any large field model is necessarily singular, and if we require that $F$ is finite for all finite values of $\phi$ we see that one cannot achieve the goal of having the cosmological solution asymptote to one in which $G=0$ always \cite{Barrowetal} without either resorting to breaking the second law of thermodynamics or having to introduce a new principle by which the Higgs field could tunnel from infinity. The asymptotic behaviour of models in which $G$ is taken to be small whilst $\phi$ grows large differs qualitatively from that in which $G$ is set by hand to zero. 

The approach of \cite{Barrowetal} 
 requires $F\rightarrow \infty$ at the origin, so that does not work. The reason for this is that in this case, $V$ is still finite at the origin, and hence the Higgs particle becomes trapped there. To illustrate this we will work in the Einstein frame. Suppose $F \approx \phi^{n}$ for $n<0$. Then, close to $\phi=0$ we find: $\chi \approx \sqrt{3}n/2 \log (\phi)$. Thus $\phi \rightarrow 0$ is $\chi \rightarrow \infty$. There is ambiguity in taking the square root of $n^2$ in this calculation. However, taking a negative square root both changes the sign of the relationship between $\chi$ and $\phi$ and flips the potential about $\chi=0$ such that the complete dynamics is insensitive to this choice. The potential in the Einstein frame is rendered differently from that in equation \ref{Ularge} :
\be U(\chi) = \frac{V}{4F^2} \rightarrow \Phi^4 e^{-\frac{4\chi}{\sqrt{3}}} \ee
This is decreasing away from the origin, hence the field would have to roll up the potential from infinity to overcome this. In terms of the original field (in the Jordan frame), this means that if $\phi$ begins at the origin it will remain there throughout the evolution. Thus we see that it is not possible that any such system (either beginning at large or small values of the field) can asymptotically become deSitter or Minkowski - it will always be singular.

\section{Further steps}\label{sec:further}
We have shown there are a number of models that can provide the kind of desired unification described in Section 1 and also be compatible with the Planck observations. 
They are large field cases, with two options 
 presented: (\ref{eq:phi_square}) and (\ref{eq:tahnh_square}). 


The agreement of these models with inflationary observations are guaranteed because during the relevant phases of inflation the dynamics is well approximated by systems already analysed in the Encyclopedia Inflationaris \cite{encyc}. One thing that should be considered carefully is the relationship between the Einstein and Jordan frames; the dynamics of inflation are usually analyzed in the Einstein frame since this where the scalar field is modelled as being minimally coupled to the (transformed) gravitational theory. However, other matter present will not have this coupling - in the Einstein frame this will appear as new couplings between the transformed field and the remaining matter. Therefore when considering reheating, for example, if one is to work in the Einstein frame, careful attention should be paid to these induced couplings. Fortunately, in the cases we consider these effects are described by the well understood $\phi^2 R$ non-minimal coupling. We have thus shown that both the question of what is the nature of the inflaton and why black holes do not dominate the entropy of the early universe can be reconciled in the same model. 

Two issues regarding the `turning off' of gravity arise. Firstly, in principle we need $1/G$ to initially be infinity to completely turn gravity off. Even in this case, we note that there is a distinct qualitative difference in dynamics between models in which this is done through taking the limit of an unbounded from above field (as presented here) and formally setting $G=0$ in the equations of motion. In practice a finite value of $1/G$ suffices to remove the entropy problem (and infinities are in any case to be avoided in real physics \cite{infinty}). What that value is depends on which detailed model is employed.  This needs further investigation, and is related to the second issue: in principle thermalisation can be achieved if the the Hubble parameter were very low to begin with, so there is time to smooth out the matter density in this weak gravity era. However in practice models do not exist where $H$ is very small initially. We therefore have to rely instead on the assumption that the universe starts off in a maximum entropy state, which will indeed be such a smooth state, instead of one chock full of black holes. 

Finally, we need an investigation of black holes for such gravity theories: What is the direct relationship between G and entropy? What is Schwarzschild solution for these cases, the horizon location, and the Hawking temperature/entropy? A naive consideration of the Bekenstein-Hawking entropy \ref{BH} indicates that (keeping matter fixed) the relationship is linear. We note that since the Schwarzschild solution is vacuum solution, it is a solution of the non-minimally coupled theory for finite $G$. However the event horizon is defined topologically as the past horizon of future null infinity - this needs modification in case in which $G$ can vary over time, and a more detailed analysis should be based on a more local concept defined by a dynamical horizon in this context \cite{DynHorizon}. 

We leave these issues for separate discussion.

\appendix
\section{the cosmological equations}\label{sec:appendix}
The cosmological equations are, the Friedmann Equation:
\be 
F\left(H^2 +\frac{F'}{F}H\dot{\phi}+\frac{k}{a^2}\right) = \rho,
\label{eq:fried1} 
\ee
the Raychaudhuri (acceleration) equation:
\be 2F(\dot{H}+H^2) = -(\rho+3P) -F'\dot{\phi}H - F''\dot{\phi}^2-F'^2 \ddot{\phi}, \label{eq:raychaud1}\ee
and the Klein-Gordon Equation:
\be  \left(1+\frac{F'}{2F}\right) (\ddot{\phi} + 3H\dot{\phi})+ V' = -\frac{F'}{F} \left(2V + \frac{\rho_m - 3P_m}{2} +(1-F'')\frac{\dot{\phi}^2}{2} \right ). \ee
The latter is equivalent to the conservation equation for the scalar field,  which in turn guarantees the consistency of (\ref{eq:fried1}) 
and (\ref{eq:raychaud1}).

\end{document}